\documentclass[prb,aps,showpacs,twocolumn,preprintnumbers,amsmath,amssymb,floatfix]{revtex4-1}
\usepackage[dvips]{graphicx}
\usepackage[latin1]{inputenc}
\usepackage{stmaryrd}
\usepackage{color}
\definecolor{dg}{rgb}{0.5,0,0} 
\usepackage{bm}      
\usepackage[normalem]{ulem}

\begin{document}
\title{The onset of magnetism peaked around $x=1/4$ in optimally electron-doped \\ LnFe$_{1-x}$Ru$_x$AsO$_{1-y}$F$_y$ (Ln = La, Nd or Sm) superconductors}
\author{S. Sanna,\,$^{1}$ P. Carretta,\,$^{1}$ R. De Renzi,\,$^{2}$ G. Prando,\,$^{1,3}$ P. Bonf\'a,\,$^{2}$ M. Mazzani,$^{2}$ G. Lamura,$^{4}$ T. Shiroka,\,$^{5,6}$  Y. Kobayashi,\,$^{7}$ M. Sato\,$^{7}$}
\affiliation{$^{1}$ Department of Physics, University of
Pavia-CNISM, I-27100 Pavia, Italy}
\affiliation{$^{2}$ Department of Physics and Earth Sciences, University of
Parma-CNISM, I-43121 Parma, Italy}
\affiliation{$^{3}$ Leibniz-Institut f\"ur Festk\"orper- und Werkstoffforschung (IFW) Dresden, D-01171 Dresden, Germany}
\affiliation{$^{4}$ CNR-SPIN and Universit\'a di Genova, via Dodecaneso 33, I-16146 Genova, Italy}
\affiliation{$^{5}$ Laboratorium f\"ur Festk\"orperphysik, ETH-H\"onggerberg, CH-8093 Z\"urich, Switzerland}
\affiliation{$^{6}$ Paul Scherrer Institut, CH-5232 Villigen PSI, Switzerland}
\affiliation{$^{7}$ Department of Physics, Division of Material Science, Nagoya University, Furo-cho, Chikusa-ku, Nagoya 464-8602, Japan}
\widetext

\date{\today}

\begin{abstract}
The appearance of static magnetism, nanoscopically coexisting with
superconductivity, is shown to be a general feature of optimally electron-doped
LnFe$_{1-x}$Ru$_x$AsO$_{1-y}$F$_y$ superconductor (Ln - lanthanide ion) upon isovalent substitution of Fe by Ru. The magnetic
ordering temperature $T_{\mathrm{N}}$ and the magnitude of the internal field display a dome-like
dependence on $x$, peaked around $x=1/4$, with higher $T_N$ values for those materials characterized by a larger $z$
cell coordinate of As. Remarkably, the latter are also those with the highest
superconducting transition temperature ($T_c$) for $x=0$. The reduction of $T_c(x)$ is found to be significant in the $x$ region of the phase diagram
where the static magnetism develops. Upon increasing the Ru content superconductivity eventually disappears, but only at $x\simeq 0.6$.

\end{abstract}

\pacs{74.70.Xa, 74.62.Dh, 76.75.+i, 74.25.Ha}
\maketitle


\section{\label{sec:intro}Introduction}
The proximity of the magnetic and superconducting ground states is
a common aspect of several strongly correlated electron systems,
ranging from the heavy fermions, to the organic materials and to
the high-temperature superconductors.\cite{Capone} Also in the
iron-based materials\cite{Hosono} superconductivity emerges close
to the disruption of a static magnetic
order.\cite{Zhao,Luetkens,Sanna2009,Drew,Shiroka} Many studies on
the transition from the magnetic to the superconducting
ground-state have been carried out in these materials, either by
varying the electron doping or by applying a high hydrostatic
pressure.\cite{Luetkens, Sanna2009, Drew, Shiroka, Maeter2013, Khasanov2011,
Julien2008, Lang2010, Wiesenmayer2011, Shermadini2012, Texier2012,
Bendele2012} In several compounds of the LnFeAsO$_{1-y}$F$_y$
family (referred to as Ln1111), with Ln a lanthanide ion, evidence
for a nanoscopic coexistence of the superconducting and magnetic states has
emerged,\cite{Sanna2009, Drew, Shiroka, Maeter2013} similarly to other Fe-based compounds. \cite{Julien2008,Wiesenmayer2011,
Shermadini2012, Texier2012, Bendele2012} The stability
of these two ground states can be investigated by perturbing the
system with, for example, a chemical substitution. In this respect
the effect of the Ru-for-Fe isovalent diamagnetic substitution is
particularly interesting. In the $y=0$ Ln1111 case this substitution
leads to a progressive dilution of the magnetic lattice\cite{McGuire2009, Yiu2012, Bonf2012} and, eventually, to
the disappearance of the magnetic order for $x\rightarrow x_c\simeq 0.6$, which is considered to be the percolation threshold for the $J_1-J_2$
localized spin system.\cite{Bonf2012}

In the optimally F-doped Ln1111 superconductor, Ru substitution leads to the
progressive reduction of the superconducting transition
temperature $T_c$ [\onlinecite{Tropeano2010},\onlinecite{Satomi}] and to its complete suppression at a Ru concentration close to $x_c$.\cite{Sanna2011}
In optimally F-doped Sm1111,
besides diluting the Fe magnetic lattice, Ru has another remarkable
effect: it induces a frozen short-range (SR) magnetic order  which
coexists nanoscopically with superconductivity albeit with reduced $T_c$ values.\cite{Sanna2011}

In this work we show that the appearance of static SR
magnetic order, nanoscopically coexisting with superconductivity,
is a common feature of Ru-doped
LnFe$_{1-x}$Ru$_x$AsO$_{1-y}$F$_{y}$ (hereafter Ln11Ru11) with: Ln = Sm, Nd or La, and with a F content close to optimal doping.
The magnetic dome is peaked around the Ru concentration $x=1/4$ and
its extension increases upon decreasing the size of
the Ln ion (since the latter is correlated with the cell
coordinate of As, $z_{As}$,  hereafter different Ln shall also be
identified by their $z$ values). On the other hand, the
superconducting transition temperature $T_c$ drops at low
Ru content, more abruptly for Ln = Sm, Nd and only marginally for
La, at the same Ru content that marks the appearance of the static SR
magnetic order at $T_{\mathrm{N}}$.

\section{\label{sec:exp_details}Experimental details}
To investigate the influence of Ru substitution on the
magnetic and superconducting properties of
LnFe$_{1-x}$Ru$_x$AsO$_{1-y}$F$_{y}$, zero- (ZF) and
longitudinal-field (LF) muon-spin spectroscopy ($\mu$SR)
experiments were performed in  powder samples with Ln = La, Nd,
prepared as reported in Ref.~\onlinecite{Satomi}. The $\mu$SR
experiments have been performed at the Paul Scherrer Institute on
the GPS and DOLLY spectrometers. ZF experiments can detect the
presence of spontaneous magnetic ordering, also in case of
short-range order. \cite{musr, Shiroka} LF experiments, instead, reveal the static
or dynamic nature of the magnetic state. \cite{musr} All the samples were
optimally electron-doped with a nominal F content $y=0.15$ and $0.11$  for Ln=Sm
and for Ln=La, Nd respectively. Here we compare
the Ln=La and Nd cases with our published Sm data.\cite{Sanna2011} The three families display
$T_c= 28$ K, 47 K, 52 K for $x=0$, respectively.

\section{\label{sec:musr} Results}

\subsection{\label{sec:musr} Detection of static magnetic order}
\begin{figure} 
\vspace{-0.5cm}
\includegraphics[width=0.5\textwidth,angle=0]{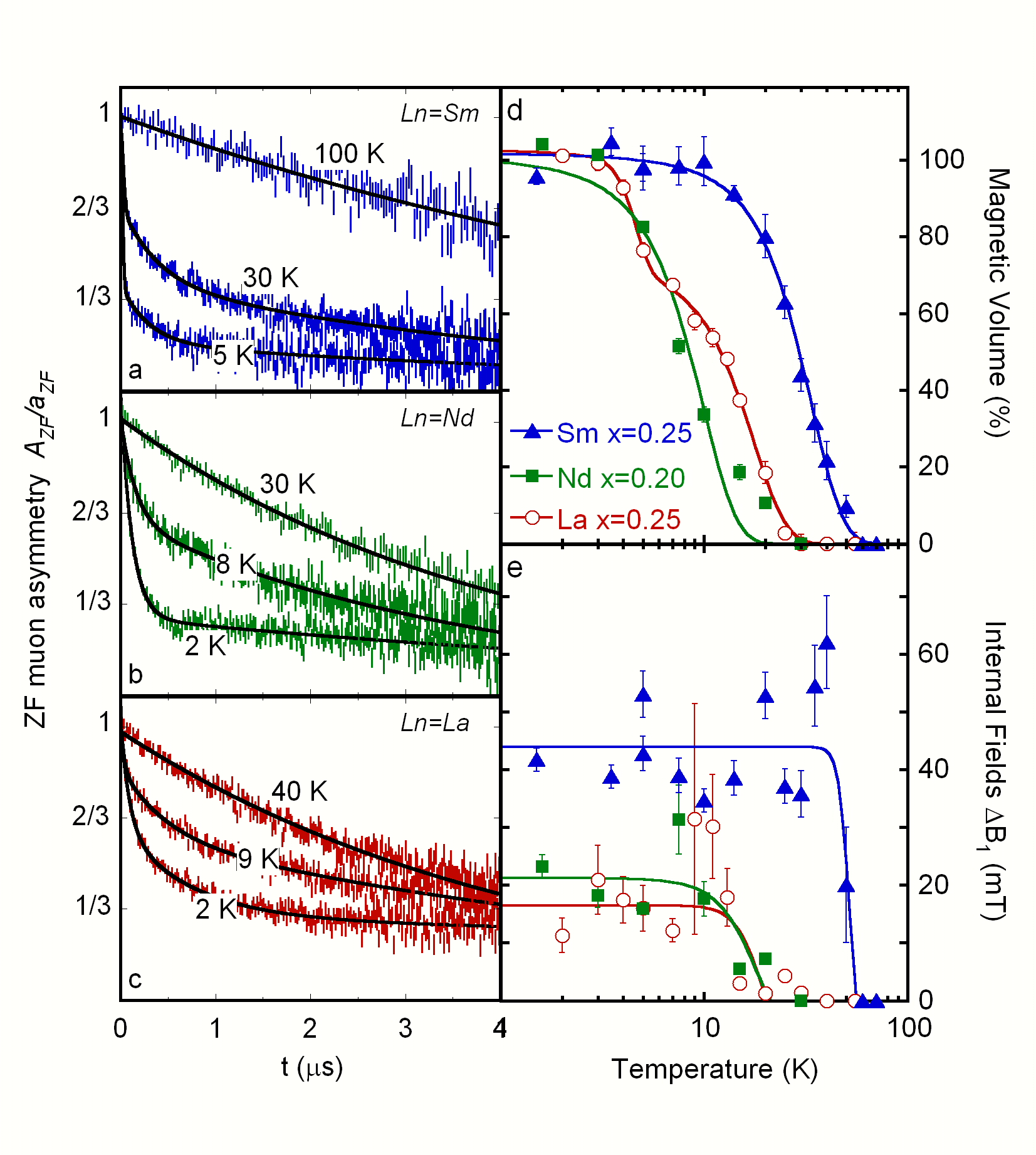}%
\caption{(Color online) ZF-$\mu$SR in optimally F-doped LnFe$_{1-x}$Ru$_x$AsO$_{1-y}$F$_{y}$ at $x\sim0.25$, with a nominal F content $y=0.15$ and $0.11$  for Ln=Sm and for Ln=La, Nd respectively.
Time dependence of the normalized muon asymmetry for Ln = Sm (a), Nd (b) and La (c).
The lines represent the best fits according to Eq.~(\ref{eq:fit_func}). Panels (d) and (e) show the volume fraction $V_{\mathrm{mag}}$, where muons detect an
internal magnetic field, and the root mean-square value of the internal field at the muon site $\Delta B_1$, respectively, as a function of temperature (see text for details).}
\label{zfmu}
\end{figure}

For each one of the Ln families under investigation, a few representative time-dependent ZF-$\mu$SR
asymmetries curves are shown in
the left panels of Fig.~\ref{zfmu}, together with the best fit to the
sum of a longitudinal and a transverse component as:

\begin{equation}
\label{eq:fit_func}
\frac{A_{\mathrm{ZF}}(t)}{a_{\mathrm{ZF}}} = \sum_{i=1,2} \left( w_{L_i} \, e^{-\lambda_it} + w_{T_i} \, e^{-\sigma^2_i t^2/2} \right)
\end{equation}

where $\sum_{i}(w_{L_i}+w_{T_i})=1$, $a_{\mathrm{ZF}}$  is the total muon signal amplitude, calibrated at
high temperature in the paramagnetic phase, $w_L$ and $w_T$ are
the weights of the transverse and longitudinal terms,
respectively, and $\lambda$ and $\sigma$ are the corresponding decay
rates. The transverse term is an overdamped precession, due to a static mean
internal field $\overline{B}$ comparable to the square-root of the second
moment $\Delta B=\sigma/\gamma$ (with $\gamma/2\pi=135.5$ MHz/T the
muon gyromagnetic ratio). Indeed, LF measurements show that for all the Ln
families an external field of the order of $\Delta B$ is sufficient to quench the transverse
relaxation, revealing the static character of the magnetic state.
The subscript $i=1,2$, when applicable, accounts for the two known muon
stopping sites in Ln1111 compounds,\cite{Maeter2009, DeRenzi2012, Prando2013}
one from within the FeAs layers and the other
close to O${}^{2-}$ ions. The two longitudinal
components $w_{L_i}$ are well resolved only for Ln=Sm, as for the non magnetic Sm1111 case.\cite{Khasanov2008}


\begin{figure}
\vspace{-1.5cm}
\includegraphics[width=0.45\textwidth,angle=0]{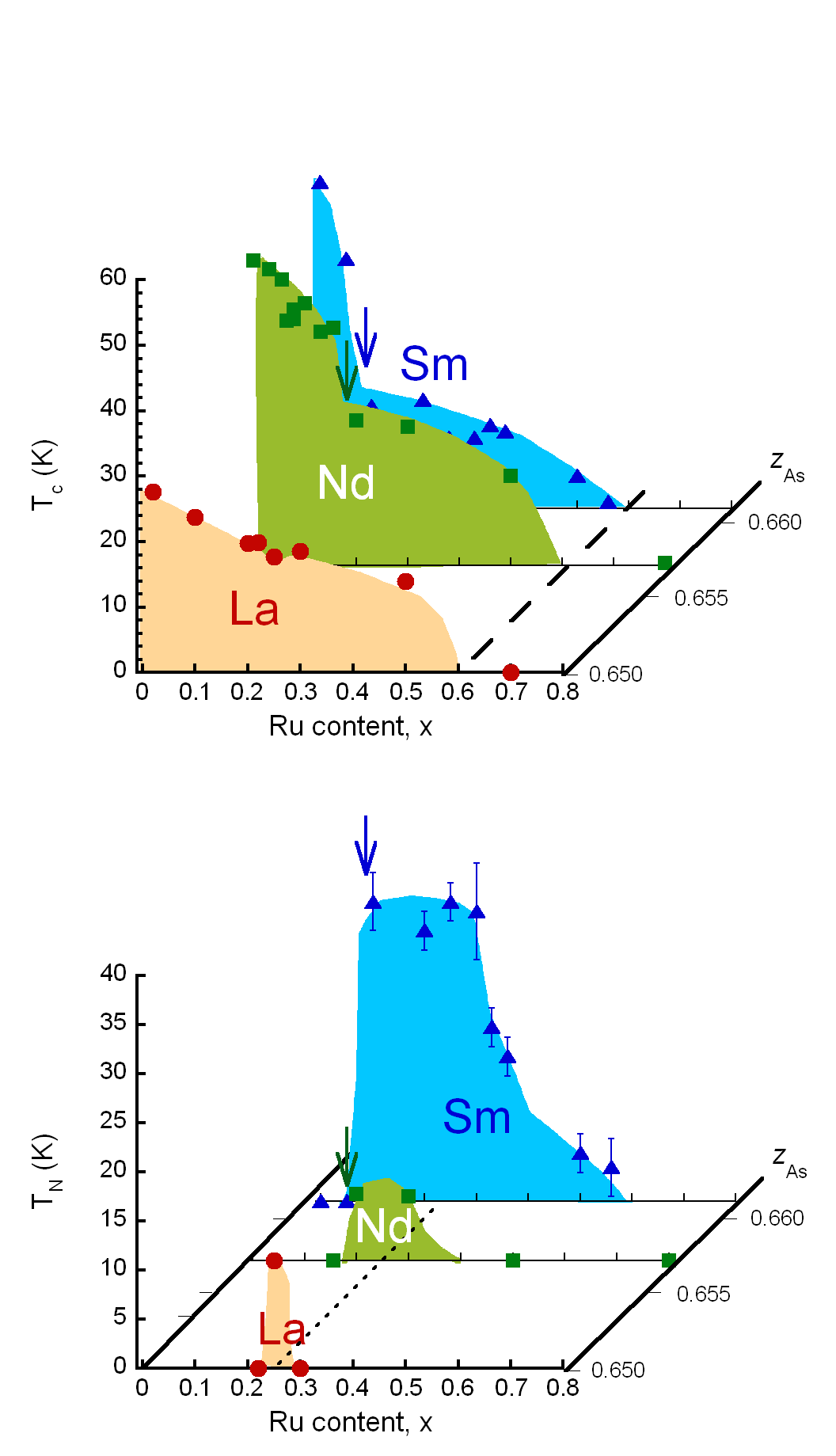}
\caption{(Color online) Transition temperatures in optimally F-doped LnFe$_{1-x}$Ru$_x$AsO$_{1-y}$F$_{y}$ (nominal $y=0.15$ and $0.11$  for Ln=Sm and for Ln=La, Nd respectively), as a function of the Ru content $x$ and the $z_{\mathrm{As}}$ cell coordinate. Top, superconducting temperatures $T_c$ from dc magnetization measurements reported in Ref. \onlinecite{Satomi,
Sanna2011}; bottom, magnetic ordering temperatures $T_{\mathrm{N}}$ as from $\mu$SR measurements. The vertical arrows indicate the $x= 0.1$ and 0.2 Ru-substitution levels in Ln = Sm and Nd, respectively. The dashed and dotted lines mark the magnetic percolation threshold $x_c\simeq 0.6$ and the $T_{\mathrm{N}}$ peak value at $x= 1/4$, respectively.}
\label{PhDiag}
\end{figure}

The magnitude of the internal magnetic fields $B_i$ at the muon site
($\simeq \Delta B_i$ shown in Fig. 1e) is the sum of the fields generated by the static moments
surrounding the muon site, which depend on the distribution of Fe and Ru atoms at the neighboring sites.
Notice that the fast decay of the transverse
component starts well above the Ln (Sm or Nd) ordering
temperature, hence it must be related to the static Fe moments.
At $T\rightarrow 0$ K the mean value of the internal fields for Ln = Sm is about 40 mT, while it
decreases to $\sim$ 20 mT for Ln = La and Nd. These are the
typical values measured in F-doped Ln1111 compounds close to the
crossover between the magnetic and superconducting phases, of Ln=Sm
and Ce [\onlinecite{Sanna2009},\onlinecite{Sanna2010}] and La [\onlinecite{Khasanov2011},\onlinecite{Prando2013}].

For polycrystals, the volume fraction in which muons experience a
net internal field can be calculated as $V_{{\mathrm{mag}}}$ $=3\sum_{i}
w_{T_i}/2=3(1-\sum_{i} w_{L_i})/2$.\cite{Sanna2009, Sanna2010} The ordering
temperatures $T_{\mathrm{N}}$ plotted in Fig.~\ref{PhDiag}b are determined
from the condition $V_{\mathrm{mag}}(T_{\mathrm{N}}) = 0.5$. Figure \ref{zfmu}d shows
that $V_{\mathrm{mag}}=1$ at low temperature for all the three Ln11Ru11 compounds. It is important
to notice that the estimated dipolar field distribution of width $\Delta
B_1$ implies a distribution of distances $0.1\lesssim r\lesssim 2$
nm between the muon and the closest frozen Fe
moment.\cite{Sanna2009}

\subsection{\label{sec:phase diagram} The 3D phase diagram}
Our main result is displayed in Fig.~\ref{PhDiag}. The top
panel shows the behavior of the superconducting transition
temperature $T_c$ for the three Ln11Ru11 families, as determined
by DC superconducting quantum interference device (SQUID)
magnetization measurements (see Ref. \onlinecite{Satomi,
Sanna2011}). The bottom panel reports the corresponding magnetic
ordering temperatures, from $\mu$SR. A remarkable feature
is the rather pronounced suppression of $T_c$ for Ln = Sm and Nd at $x \simeq 0.10$ and 0.20, respectively.
As already shown\cite{Sanna2011}
for Ln = Sm this concentration coincides with the onset of static SR
magnetic order (see Fig.~\ref{PhDiag} bottom panel). A very similar behavior is observed also
in case of Ln=Nd at $x\simeq0.2$. $T_c$ vanishes for {\em all} the families around
$x\rightarrow 0.6$, which corresponds to the magnetic percolation threshold $x_c$
for the magnetic lattice in the undoped La1111.\cite{Bonf2012}

In addition it is interesting to note that, despite the low density of Ln=Nd points, both La and Sm data give evidence that the maximum of $T_N$ is peaked around $x=0.25$. In the bottom panel of Fig.~\ref{PhDiag} this indication is very sharp for Ln=La. For Ln=Sm, although asymmetric, $T_N$ has again a maximum around $x=0.25$.

\subsection{\label{sec:coexistence} Nanoscopic coexistence of magnetism and superconductivity}
We found that $V_{{\mathrm{mag}}}$ is close to unity below $T_{\mathrm{N}}$
for all the magnetic samples with nonzero $T_{\mathrm{N}}$.
This implies that in the absence of an applied field all ZF implanted muons detect the
presence of an ordered magnetic moment, i.e. internal fields
develop throughout the whole sample (although some muons detect
these fields from just outside a magnetic region, not farther than
a few nm). Since dc magnetometry measurements detect a sizeable
superconducting fraction,\cite{Satomi, Sanna2011} following the
same arguments used in Refs. \onlinecite{Sanna2010,Sanna2011}, it is conceivable that in all these families magnetic and
superconducting regions form an interspersed texture, owing to the
nanoscopic electronic inhomogeneities induced by the Ru distribution.

\begin{figure}
\includegraphics[width=0.5\textwidth,angle=0]{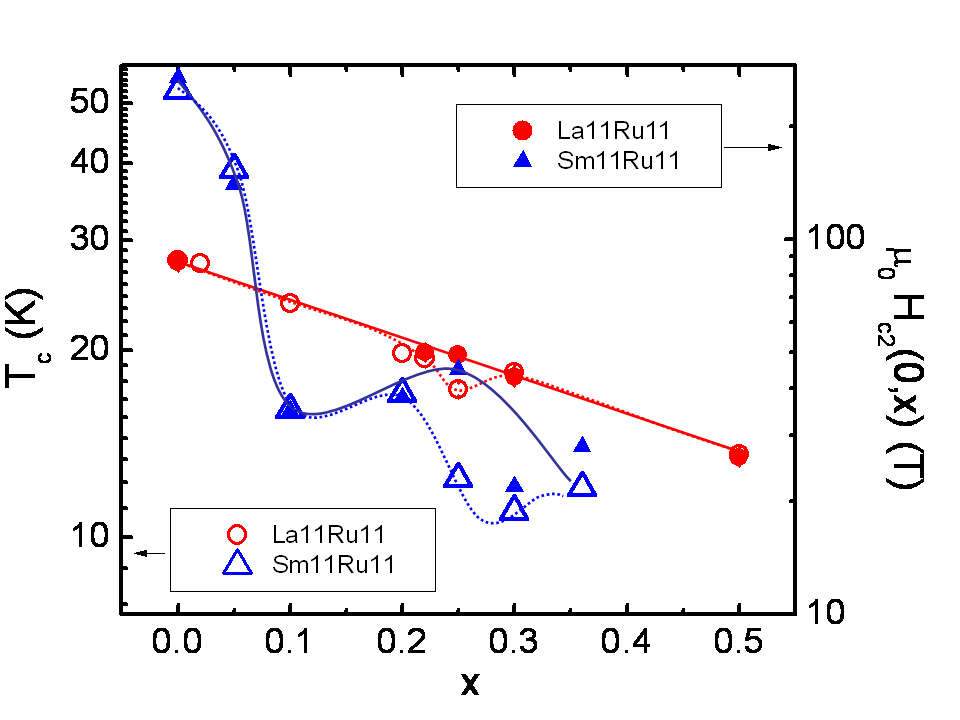}%
\caption{(Color online) Superconducting $T_c$ (open symbols, left
scale) and  $H_{c2}(T\rightarrow 0)$ (closed symbols, right
scale), vs. Ru content $x$ in
SmFe$_{1-x}$Ru$_x$AsO$_{0.85}$F$_{0.15}$ (triangles) and in
LaFe$_{1-x}$Ru$_x$AsO$_{0.89}$F$_{0.11}$ (circles). The lines are
guides to the eye. } \label{Hc2}
\end{figure}

However superconductivity may survive in such finely dispersed
regions only as long as its coherence length $\xi$ is comparable
to the average separation $r$ (few nm) among magnetic domains. In
order to roughly estimate the coherence length $\xi_x$ as a function of Ru
content $x$, the upper critical field $H_{c2}(T,x)\propto
\xi_x^{-2} $ was derived\cite{Prando2012} for Ln = La. The data
for Ln = Sm are taken from Ref.~\onlinecite{Tropeano2010}. The
value of $H_{c2}$ for $T\rightarrow 0$ was estimated from the
Werthamer-Helfand-Hohenberg relation\cite{WHH}
\begin{equation}
H_{c2}\simeq 0.7\times T_c(H=0)|dH_{c2}/dT|_{T_c(x,H=0)}\,\,\, .
\end{equation}
Although this expression tends to overestimate\cite{Hc2}
$H_{c2}(0,x)$, it can still provide the relative variation of the upper
critical field with Ru substitution. The results are shown in
Fig.~\ref{Hc2} for Ln = La (solid circles) and Sm (solid
triangles). One can then derive $\xi_x$  and find that
$\xi_{0.25}/\xi_0$ is 1.3 and 2.4 for La and Sm, respectively.
Hence, from the absolute values of $\xi_0$ reported in
Ref.~\onlinecite{Prando2012} one estimates an absolute value
$\xi_{0.25} \approx$ 3 nm for both Ln ions, namely the same order
of magnitude of the mean distance $r$ among
magnetic domains. Thus, the observation of bulk superconductivity does not conflict
with muons detecting $V_{mag} = 1$ (Fig.~\ref{zfmu}d).

The nanoscopic coexistence of the two phases in Ln11Ru11 is
reminiscent of that observed at the crossover between magnetic and
superconducting order in F-doped Sm1111 \cite{Sanna2009,Drew} and
Ce1111 \cite{Sanna2010}, but notably not in La1111. In the latter
the two order parameters are mutually exclusive at ambient
pressure \cite{Luetkens} and are observed to coexist in
mesoscopically separated regions under both external\cite{Khasanov2011}
and chemical \cite{Prando2013b} pressures. Here, the detection of nanoscopic
coexistence not only in Sm11Ru11 and Nd11Ru11, but also in La11Ru11, suggests that
the substitution of Fe with the isovalent non magnetic Ru induces a static
SR magnetic order.

\section{\label{sec:discussion}Discussion}
%

\begin{figure}
\includegraphics[width=0.45\textwidth,angle=0]{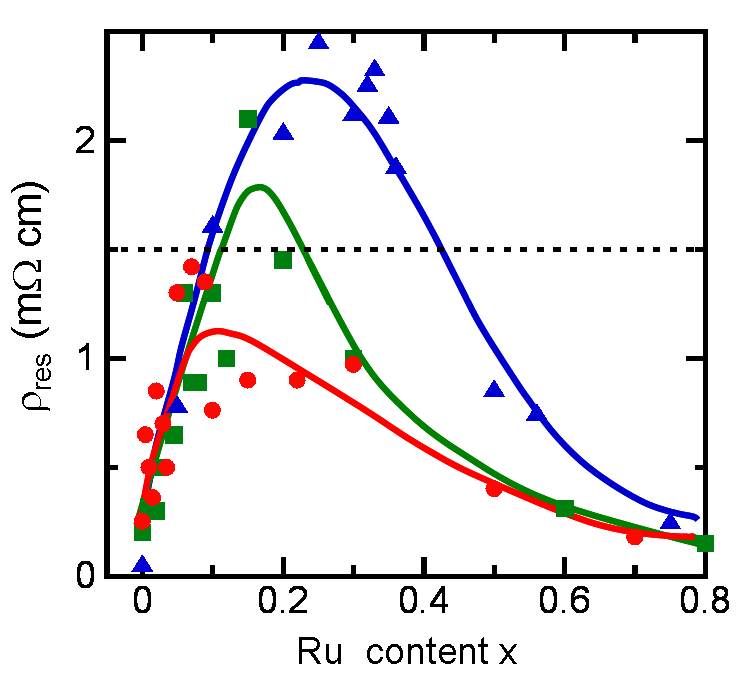}%
\caption{(Color online) Residual resistivity as a function of Ru
content for LnFe$_{1-x}$Ru$_x$AsO$_{0.89}$F$_{0.11}$ with Ln=La (circle), Nd (square) and Sm (triangle) families,
using data reported in Ref.[\onlinecite{Satomi}] and [\onlinecite{Tropeano2010}].} \label{rho}
\end{figure}

Our results show that the static SR magnetic order induced
by the isovalent and nonmagnetic Ru substitution around $x=1/4$ is
a common aspect of LnFe$_{1-x}$Ru$_x$AsO$_{1-y}$F$_y$ optimally
F-doped superconductors.

A possible mechanism which explains the observed suppression of $T_c$ and the appearance of a static
SR magnetic order is the electron localization, arising from Ru impurity scattering.
This localizing effect is well know to suppress the superconducting transition temperature \cite{deGennes}.
In addition, due to the loss of the kinetic energy of the electrons, which becomes significant as the temperature decreases,
electron localization may induce local magnetic moments on the Fe ions which eventually freeze below $T_{\mathrm{N}}$.

The above idea is supported by the $x$ dependence
of the residual resistivity, $\rho_{res}$, which for all
Ln = La, Nd \cite{Satomi} and Sm \cite{Tropeano2010} compounds
displays a behavior analogous to the one of $T_N(x)$, as shown in
Fig.~\ref{rho}. This dome-like trend indicates a competition
between the electron localization and the increase in kinetic energy
caused by the more extended Ru $d$ orbitals.
Interestingly, at least for Sm and Nd, the maximum of the residual
resistivity is close to $x=1/4$ where $T_N(x)$ is peaked.
Moreover, a direct comparison with Fig.~\ref{PhDiag}b indicates
that the magnetic state can develop only for those samples with
$\rho_{res}\gtrsim\rho_{c}=1.5$ $m\Omega cm$ (emphasized by the
dotted line), which is quite reasonable as a crossover value for Anderson localization. \cite{Sefat2006}

In a slightly different scenario, one can consider that the perturbation
generated by Ru impurity yields a staggered polarization of the magnetic moments present on the surrounding Fe sites.
This hypothesis is based on the experimental evidence for the existence of a sizeable, rapidly fluctuating
magnetic moment at the Fe site in different optimally doped Fe-based superconductors.
\cite{Vilmercati2012, Liu2012} Then, one could expect that when these moments freeze, a static SR order appears, characterized
by antiferromagnetic correlations analogous to those observed in the underdoped  Ln=Sm, Ce
[\onlinecite{Sanna2009},\onlinecite{Sanna2010}] and Ln=La
[\onlinecite{Khasanov2011}] compounds close
to the crossover between magnetic and superconducting phases. In fact, the $\mu$SR spectra shown in
Fig.~\ref{zfmu} at $x\simeq 1/4$ are very similar to those measured
at that crossover. Specifically, no oscillations are observed in the time spectra and the depolarization
rates of the transverse fractions measured here for $T\rightarrow0$, when Ln=Sm or La,
are similar to those of the F-doped Ln1111 compounds close to the crossover between magnetic and superconducting phases,
namely $\sigma_1$ is of about $60 ~\mu s^{-1}$ [9] and $20 ~\mu s^{-1}$ [14], respectively.

We notice that the appearance of static magnetism (around $x=0.1$, Ln=Sm and $x=0.2$, Ln=Nd, arrows in the top and bottom panels of
Fig.~\ref{PhDiag}) is concomitant with a marked change of the derivative $dT_c(x)/dx$. Figure \ref{Hc2} shows that
the critical field $H_{c2}(x)$ generally follows the same trend of
$T_c(x)$ for both Sm and La, with the notable exception of the
$x\simeq0.25$ compositions, where $T_c$ is more drastically depressed
in both families. The effect is more sizeable when the SR magnetic
order is stronger (Ln=Sm, Fig.~\ref{PhDiag}). The comparison of Sm
and La at $x=0.1$ indicates that the onset of magnetic
order in Sm11Ru11 depresses $T_c$ well below the value of the
corresponding La11Ru11, where the static SR magnetic order is absent. This behavior
indicates that the {\em static} magnetism and superconductivity in
1111 do strongly compete. In other words, $T_c$ seems to be reduced
by the renormalization of the spectum of the spin fluctuations, induced by
the onset of the static SR order, suggesting that superconductivity
is driven by a spin fluctuation mechanism. This observation
appears to be in agreement with recent point-contact Andreev-reflection measurements\cite{Daghero2012}
performed on the same set of Sm11Ru11 samples, which indicate a
progressive decrease of the boson energy when static magnetism
appears.

The static magnetic order
is more extended and accompanied by larger internal fields $\Delta
B_1$ in those compounds where the superconducting $T_c$ for the Ru free
$x=0$ composition is higher. Indeed, both $\Delta B_1$ and $T_c$
increase from La to Sm, together with the $z$ cell coordinate of
As,\cite{Iadecola2009, Johnston2010}(oblique axis in Fig.~\ref{PhDiag}) that is very little Ru
dependent \cite{Tropeano2010, Iadecola2012}. This trend is in
agreement with Landau free energy derivation, based on density
functional calculations in the local density approximation,
showing that the magnetic ground-state in Ln1111 compounds gets
progressively more stable as $z$ increases.\cite{Jose} The same
calculations suggest that the $z_{\mathrm{As}}$ coordinate may effectively
tune the approach to a Quantum Tricritical Point (QTP) where an
enhancement of the superconducting pairing may occur.

On the other hand, it should be pointed out that the strong competition between the superconductivity and
the static SR magnetic order can be understood even when the pairing
mechanism is not related to the spin-fluctuations. Actually, the $s_\pm$
symmetry expected in case of spin-fluctuation mechanism can hardly explain the very small initial $T_c$-suppression rate $|dT_c/dx|_{x\rightarrow 0}$
observed in these systems, \cite{Onari2009,
Sato2010, Sato2012} unless the intra-band impurity scattering is much larger than the inter-band one.\cite{Chubukov1,Chubukov2,Wang2013}
On this point, it is worth mentioning that theories which predict a possible role of the orbital fluctuations
\cite{Kontani2010,Onari2012,Yanagi2010} predict an $s_{++}$ symmetry of the order parameter, which can explain the small values of
$|dT_c/dx|_{x\rightarrow 0}$ observed here.

\section{Conclusion}

In conclusion, the phase diagram of Ru-doped LnFeAsO$_{1-y}$F$_y$ with optimum value of $y$
was outlined for different Ln ions. It was shown that the appearance of static magnetism induced
by the nonmagnetic isovalent Ru substitution around $x=1/4$ is
a common aspect of LnFe$_{1-x}$Ru$_x$AsO$_{1-y}$F$_y$ optimally
F-doped superconductors. The onset of the magnetism is concomitant to a sizeable weakening of
the superconducting state in the $x$ region where the residual resistivity shows a peak, namely where
the effects of the electron localization are most significant. The stronger the static magnetism
induced by Ru substitution, the more significant is the degradation of the superconducting state, which
is definitely suppressed only for $x\rightarrow x_c\simeq 0.6$. In
addition, it was shown that the magnitude of the transition temperature $T_{\mathrm{N}}$
and the $x$ extension of the magnetic phase appear to progressively vanish as one moves in the
$x-z_{\mathrm{As}}$ plane towards lower $z_{\mathrm{As}}$ values while $x$ is kept at $\sim 1/4$.
These results were discussed in the framework of different superconducting pairing mechanisms.

\begin{acknowledgments}
This work was performed at the Swiss Muon Source S$\mu$S,
Paul Scherrer Institut (PSI, Switzerland) and was partially supported
by Fondazione Cariplo (research grant no.\ 2011-0266) and by MIUR-PRIN grant
2008XWLWF9-04. T.S.\ acknowledges support from the Schweizer Nationalfonds (SNF)
and the NCCR program MaNEP. G.P.\ acknowledges support from the Leibniz-Deutscher Akademischer Austauschdienst (DAAD) Post-Doc
Fellowship Program. The assistance by Alex Amato and Hubertus Luetkens during the $\mu$SR measurements
at PSI is gratefully acknowledged.
\end{acknowledgments}



\end{document}